\documentclass[12pt]{iopart}

\usepackage{graphicx,iopams,color}

\begin{document}

\title{Attenuation of transcriptional bursting in mRNA transport}

\author{Li-ping Xiong$^1$, Yu-qiang Ma$^1$ and Lei-han Tang$^2$}

\address{$^1$ National Laboratory of Solid State Microstructures,
Nanjing University, Nanjing 210093, China

$^2$ Department of Physics, Hong Kong Baptist University, Kowloon
Tong, Hong Kong} \ead{lhtang@hkbu.edu.hk}
\begin{abstract}
Due to the stochastic nature of biochemical processes,
the copy number of any given type of molecule inside a
living cell often exhibits large temporal fluctuations.
Here, we develop analytic methods to investigate how the noise
arising from a bursting input is reshaped by a transport reaction
which is either linear or of the Michaelis-Menten type.
A slow transport rate smoothes out fluctuations at the output end
and minimizes the impact of bursting on the downstream cellular activities.
In the context of gene expression in eukaryotic cells, our results
indicate that transcriptional bursting can be substantially attenuated
by the transport of mRNA from nucleus to cytoplasm.
Saturation of the transport mediators or
nuclear pores contributes further to the noise reduction.
We suggest that the mRNA transport should be taken into account in the
interpretation of relevant experimental data on transcriptional bursting.

\end{abstract}

\maketitle

\section{Introduction}

Molecular binding and chemical modifications underlying
intracellular processes are intrinsically stochastic.
They give rise to temporal fluctuations and cell-to-cell variations
in the number of molecules of any given type, mask genuine signals
and responses, and generally contribute to the phenotypic diversity
in a population of genetically identical
individuals\cite{eukaryotic1Oshea,cellfateDubnau}.
Various characteristics of such noise have been under intense
quantitative study in the past few years
\cite{negativeSerrano,networkOudenaarden}.
One of the focal points of
the discussion is how the noise propagates along a biological pathway.
It has been shown that in cases where the dynamics of the upstream
molecules is not affected by the downstream processes
(e.g., in gene transcription and translation),
a ``noise addition rule'' generally applies, i.e., each process in the
pathway contributes to the overall noise strength in a statistically
independent fashion\cite{additionPaulsson,memoryPaulsson}.
Modifications to this rule in molecular circuits with feedbacks or
``detection'' capabilities as in signalling have been examined
by T\u{a}nase-Nicola and his colleagues\cite{correlationWolde}.

In metabolic pathways and other transport processes, however,
the upstream molecule is passed on to the downstream pool in a modified form.
The conservation of mass and/or number of molecules in the reaction
sets new rules on noise propagation.
In a recent work, Levine and Hwa considered this problem in
the steady state of a metabolic network\cite{metabolicHwa}.
Their results show that fluctuations in the number of
intermediate metabolites are generally uncorrelated to each other,
upstream or downstream along a pathway or across branches.
This, of course, does not exclude dynamic correlations which can be
quite nontrivial in driven processes\cite{kpz}.
The full dynamic description of stochastic transport through a network is
a very challenging task which has not been fully solved even for a linear
pathway\cite{schuetz}. A case of interest in the present context is busty
input, where the molecules to be transported are produced in batches
separated by long silent intervals. Well-known examples of such behavior
include the transcriptional and translational
bursting\cite{eukaryotic1Oshea,reviewCollins} and vesicular
transport\cite{endocytosisStenmark}.
In eukaryotic gene expression, a large number of mRNAs are produced
over a short period of time and followed by a long silent period as
a result of the chromatin remodeling.
In prokaryotes, burst of protein copy number can occur as
the result of short lifetime of an mRNA transcript.
Nutrient uptake in endocytosis can also be
viewed as a burst event: each endocytic vesicle transports and
releases a large number of extracellular molecules to the target site.

In this paper, we examine the attenuation of bursting noise
in a two-compartment model with a stochastic transport channel.
The model can be viewed as an improvement over the one-compartment
model used previously to analyze the mRNA copy number fluctuations
in single-molecule experiments carried out by Raj
{\it et al.}\cite{mamalianTyagi}
on mammalian cell gene expression.
The experiments show that the number of mRNA
transcripts produced in a single burst event ranges from a few
tens to hundreds. It was argued that such large bursts, if unattenuated,
could harm the progression of normal cellular
activity\cite{minimizationEisen} due to their large perturbations
to the cytoplasmic mRNA and protein levels.
A suggestion was made by Raj {\it et al.} that the latter could be avoided
if the overall protein copy number is kept high by a low protein
degradation rate. Our analysis here points to a second possibility:
a slow nuclear mRNA processing and export process can also
attenuate mRNA bursting and minimize its impact on the downstream
protein population.

Supporting this view, an earlier kinetic
study of mammalian cell mRNA splicing and nuclear transport has
shown that the nuclear dwelling (or retention) time of an mRNA
molecule can be comparable to its lifetime in the cytoplasm
\cite{kineticsDautry}. This is evidenced in the time required to
reach the respective steady state levels for the mRNAs residing in
the nucleus and in the cytoplasm, which were measured separately in
the experiment. Consistent with this observation, both studies also
concluded that, on average, about $10$-$40$\% of mRNA are retained
in the nucleus. In this respect, the one-compartment model
of Raj {\it et al.}, which considers only the total
mRNA copy number in a cell, overestimates
fluctuations in the actual number of mRNAs in the cytoplasm that
participate in the translation process.

To establish a reasonable model, let us first examine the typical fate
of a single mRNA: the mRNA is synthesized in bursts at the transcription
site and almost simultaneously processed into an mRNA-protein (mRNP)
complex\cite{unifiedReinberg,complexKataoka}; the mRNP complex
diffuses inside the
nucleus\cite{nucleiSinger,transport2Tyagi,traffickingMisteli},
eventually reaches one of the nuclear pores and exits with
the help of export mediators\cite{transport1Adam,transport3Scarcelli};
the mRNA degrades in the cytoplasm. Therefore, our model includes
three processes: the transcriptional bursting in the nucleus,
the mRNA transport, and the mRNA decay in the cytoplasm. We shall
assume that the mRNA transport to the cytoplasm is much slower than
the diffusion process, so that the spatial inhomogeneity
of the mRNA molecules in the two compartments can be ignored\cite{note1}.

Both linear and nonlinear transport are considered.
In linear transport, the export mediators and nuclear pores
are abundant as compared to the transported mRNA. Consequently, the export
events are independent of each other.
Nonlinear transport corresponds to a situation where queuing of nuclear
mRNA takes place due to a limited number of transport mediators or
nuclear pores. Its mathematical description is identical to
that of the Michaelis-Menten (MM) reaction in enzyme kinetics.

The stochastic bursting and transport processes can be described
by the chemical master equation governing the time-dependent distribution
of mRNA copy numbers in the two compartments.
In the case of linear transport, exact expressions for the
copy number fluctuations in a steady-state situation are obtained.
The MM transport in the weak noise limit can be treated
using the linear noise approximation (LNA)\cite{stochasticKampen}.
A novel independent burst approximation (IBA) is introduced to
treat the MM transport in the strong fluctuations regime.
We also perform stochastic simulations to check
the accuracy of the analytic expressions.

The main results of our calculation are summarized in Sec. 3.
The noise strength of cytoplasmic mRNA is expressed in terms of
the average burst size and the ratio of the mean nuclear and
cytoplasmic mRNA copy numbers, which are measurable in experiments.
Quite generally, the extent of burst attenuation is governed by the
rate of transport.
The slower the mRNA transport, the smaller is the noise in the
cytoplasmic mRNA number. In the case of the MM transport,
the saturation effect of transport mediators or nuclear pores
further reduces mRNA copy number fluctuations in the cytoplasm.
Based on these findings, we suggest a revision of the parameters
estimated by Raj {\it et al.} for the bursting and decay dynamics of mRNA
in their experiments.

\section{Methods}

The transport event involves two compartments, usually with
different volumes which result in different concentrations even
when the number of molecules is the same. To step aside this problem
and to be more clear, we measure the amount of mRNA in copy number
rather than concentration, and define all the reaction rates
mesoscopically(by scaling the macroscopic counterparts with
volumes). Such a treatment (we will not explicitly refer to specific
units for variables and parameters in the following calculations)
also facilitates the application of chemical master equations.
Then we can start safely by introducing the methods for the
simplest case: linear transport.

\subsection{Linear transport}

The burst of mRNA arises from the transitions between active gene
$A$ and inactive gene $I$, while each active gene produces an amount
of nuclear mRNA $M_{\rm n}$. To faithfully describe the bursting, we
write the relevant reactions, as often done in many previous works
\cite{eukaryotic1Oshea,mamalianTyagi}, as:
\begin{equation}
I \stackrel{\lambda}{\longrightarrow} A,\quad   A
\stackrel{\gamma}{\longrightarrow} I, \quad A
\stackrel{\mu}{\longrightarrow} A + M_{\rm n}.
\label{gene_activation}
\end{equation}
Here $\lambda$ and $\gamma$ are the rates of gene activation and
inactivation respectively, and $\mu$ is the transcription rate. 
Using $M_{\rm c}$ to denote the cytoplasmic mRNA, 
we define the linear transport and cytoplasmic decay of mRNA as: 
\begin{equation}
M_{\rm n} \stackrel{k}{\longrightarrow} M_{\rm c},\quad   M_{\rm c}
\stackrel{\delta}{\longrightarrow} \O.
\label{linear-transport-decay}
\end{equation}
Here $k$ is the transport rate, and $\delta$ is the degradation
rate of cytoplasmic mRNA. The decay of mRNA in nucleus is ignored
\cite{kineticsDautry}.

At a given time $t$, the gene can be in either active or inactive state,
with the probabilities given by $P_{A}(m_{\rm n},m_{\rm c},t)$ and
$P_{I}(m_{\rm n},m_{\rm c},t)$, respectively. The probabilities depend
on the copy number of mRNA in nucleus $m_{\rm n}$ and in cytoplasm
$m_{\rm c}$. The time evolution of the two probabilities are governed
by the chemical master equations:
\begin{eqnarray}
\label{mastequation1} \frac{dP_{A}(m_{\rm n},m_{\rm c},t)}{dt}&=&\lambda
P_{I}(m_{\rm n},m_{\rm c},t)-\gamma
P_{A}(m_{\rm n},m_{\rm c},t)\nonumber\\
&&+\mu(\varepsilon_{\rm n}^{-1}-1)P_{A}(m_{\rm n},m_{\rm c},t)\nonumber\\
&&+k (\varepsilon_{\rm n}\varepsilon_{\rm c}^{-1}-1)m_{\rm n}
P_{A}(m_{\rm n},m_{\rm c},t)\nonumber\\
&&+\delta (\varepsilon_{\rm c}-1)m_{\rm c}P_{A}(m_{\rm n},m_{\rm c},t),\\
\label{mastequation2}\frac{dP_{I}(m_{\rm n},m_{\rm c},t)}{dt}&=&\gamma
P_{A}(m_{\rm n},m_{\rm c},t)-\lambda
P_{I}(m_{\rm n},m_{\rm c},t)\nonumber\\
&&+k (\varepsilon_{\rm n}\varepsilon_{\rm c}^{-1}-1)m_{\rm n}
P_{I}(m_{\rm n},m_{\rm c},t)\nonumber\\
&&+\delta (\varepsilon_{\rm c}-1)m_{\rm c}P_{I}(m_{\rm n},m_{\rm c},t),
\end{eqnarray}
where $\varepsilon$ is the step operator defined by its effect on
arbitrary functions of $m_{\rm n}$ and $m_{\rm c}$:
$\varepsilon_{\rm n}^{\pm 1}f(m_{\rm n},m_{\rm c},t)
=f(m_{\rm n}\pm 1,m_{\rm c},t)$ and
$\varepsilon_{\rm c}^{\pm 1}f(m_{\rm n},m_{\rm c},t)
=f(m_{\rm n},m_{\rm c}\pm 1,t)$.

The first moments or mean values of $m_{\rm n}$ and $m_{\rm c}$ can be
obtained by multiplying equations \eref{mastequation1} and
\eref{mastequation2} by $m_{\rm n}$ and $m_{\rm c}$ in turn, and summing
over all $m_{\rm n}$, $m_{\rm c}$ and gene states:
\begin{eqnarray}
\label{mean}
\frac{d\langle m_{\rm n}\rangle}{dt}&=&\mu p_{A}-k\langle m_{\rm n}\rangle,
\label{linear-mean}\\
\frac{d\langle m_{\rm c}\rangle}{dt}&=&k\langle m_{\rm n}\rangle-\delta
\langle m_{\rm c}\rangle.
\label{linear-variance}
\end{eqnarray}
Here $\langle\cdot\rangle$ denotes average over the distribution, and
$p_{A}(t)\equiv\sum_{m_{\rm n},m_{\rm c}}P_A(m_{\rm n},m_{\rm c},t)$
is the probability that
the gene is in the active state. The following equation is easily
seen from the gene activation dynamics (\ref{gene_activation}):
\begin{equation}
dp_{A}/dt=\lambda (1-p_{A})-\gamma p_{A}.
\label{activation_dynamics}
\end{equation}
Equations (\ref{linear-mean})-(\ref{activation_dynamics})
are equivalent to the macroscopic rate equations given
by the mass-action law due to the linearity of the process. Setting
the right-hand-side of these equations to zero, we obtain the steady-state
relations for the average mRNA flux:
$J=\mu p_A^*=\mu\lambda/(\lambda+\gamma)
=k\langle m_{\rm n}\rangle=\delta \langle m_{\rm c}\rangle.$

In the following discussion we will focus on
the burst limit where $\mu$ and $\gamma$ are significantly
larger than all other reaction rates. The steady-state
probability $p_A^*$ goes to zero but
the mRNA synthesis rate $J=\mu p_A^*\simeq \lambda(\mu/\gamma)$
remains finite. Gene activation in this case follows a Poisson process
at a rate $\lambda$, but the number of mRNA copies $b$ produced in each
burst event is a random variable that satisfies the geometric
distribution: $G (b)= (\mu/\gamma)^{b} (1+\mu/\gamma)^{-b-1}$
\cite{transcriptionalKondev}.
In terms of the mean mRNA copy number produced
in each burst, $\langle b\rangle = \mu/\gamma$, we have
\begin{equation}
J=\lambda\langle b\rangle =k\langle
m_{\rm n}\rangle=\delta \langle m_{\rm c}\rangle.
\label{linear-flux}
\end{equation}
A similar procedure as above yields the second moments of mRNA copy
numbers in the steady state:
\begin{eqnarray}
\label{mn2apr}\langle m_{\rm n}^{2}\rangle&=&\langle
m_{\rm n}\rangle^{2}+ (\langle b\rangle+1)\langle m_{\rm n}\rangle,\\
\label{mnmcapr}\langle m_{\rm n}m_{\rm c}\rangle&=&\langle
m_{\rm n}\rangle\langle m_{\rm c}\rangle+\langle b\rangle\frac{\langle
m_{\rm n}\rangle \langle m_{\rm c}\rangle}{\langle m_{\rm n}\rangle+\langle
m_{\rm c}\rangle},\\
\label{mc2apr}\langle m_{\rm c}^{2}\rangle &=&\langle
m_{\rm c}\rangle^{2}+ (\langle b\rangle+1)\langle m_{\rm c}\rangle-\langle
b\rangle\frac{\langle m_{\rm n}\rangle \langle m_{\rm c}\rangle}{\langle
m_{\rm n}\rangle+\langle
m_{\rm c}\rangle}.
\end{eqnarray}

It is customary to measure temporal variations of population size in a
stationary process using the noise strength (also known as the Fano factor),
defined as the variance over average \cite{additionPaulsson,reviewCollins}:
\begin{eqnarray}
\label{mnff}
\frac{\sigma_{m_{\rm n}}^{2}}{\langle m_{\rm n}\rangle}
&=&\langle b\rangle+1,\\
\label{mcff}\frac{\sigma_{m_{\rm c}}^{2}}{\langle
m_{\rm c}\rangle}&=&\langle b\rangle+1-\langle b\rangle\frac{\langle
m_{\rm n}\rangle}{\langle m_{\rm n}\rangle+\langle m_{\rm c}\rangle}.
\end{eqnarray}
For the Poissonian fluctuation arising from the simplest case where
molecules are produced one by one with a constant probability and
degraded linearly, the noise
strength is unity\cite{focusingPNASEhrenberg,focusingPRLEhrenberg}.
When the synthesis is burst-like and the degradation is still linear, the
noise strength becomes $\langle b\rangle+1$, which is much larger
than the Poissonian fluctuation \cite{burstOudenaarden}. This is the
case for the nuclear mRNA population [equation \eref{mnff}], and for
the mRNA without transport, as in the prokaryotic cells. Equation
\eref{mcff}, on the other hand, shows that although the transport
event follows a random process, the noise strength of the
cytoplasmic mRNA that propagates directly to protein noise, is
actually reduced. The amount of reduction is controlled by the ratio
$\langle m_{\rm n}\rangle/(\langle m_{\rm n}\rangle+\langle m_{\rm c}\rangle)$,
which increases with decreasing transport rate $k$.

\subsection{Michaelis-Menten-type transport}

Accumulation of the mRNAs in the nucleus may lead to a saturation
effect that changes the transport dynamics when the
number of export mediators or nuclear pores becomes limiting.
This prompts us to study a more general mechanism of
transport that takes the transport capacity into account.
The resulting process
can be cast in the form of the well-known Michaelis-Menten model.

To simplify the discussion, we consider here only
one source of constraint, say the limited number of one kind of
export mediator denoted by $E$, and treat the rest of the
export mediators (including the nuclear pores) in the
process to be non-rate-limiting.
Thus, the transport process of mRNA can be described as:
\begin{equation}
M_{\rm n}+E {\scriptstyle k_{1}  \atop
\stackrel{\displaystyle\rightleftharpoons}{\scriptstyle k_{2}}} EM_{\rm n}
\stackrel{k_{3}}{\longrightarrow} E+M_{\rm c},
\label{MM-rates}
\end{equation}
where $k_{1}$ and
$k_{2}$ are the binding and unbinding rates respectively, and
$k_{3}$ is the export rate. This is similar to the MM model for
enzymatic reaction.

The analysis presented below is based on the ``fast equilibration''
approximation, in which $EM_{\rm n}$ is treated as a transition state
rather than an accumulation point in the mRNA export.
For this to be true, the lifetime of the complex $EM_{\rm n}$ should
be significantly shorter than the total nuclear dwelling time,
i.e., either the Michaelis constant $K_{\rm m}= (k_{2}+k_{3})/k_{1}$
is much greater than one, or $k_3$ is much smaller than the other two rates.
Under this assumption, the $EM_{\rm n}$ population remains in
quasiequilibrium with the nuclear population $m_{\rm n}$ which varies
on a much slower time scale as compared to the decomposition time of
the complex $EM_{\rm n}$.

Due to the small copy number of the nuclear mRNA and the transporter $E$,
we distinguish the free $M_{\rm n}$ from the bound ones, and define the
total number of nuclear mRNA as $m_{\rm n}=m_{\rm nf}+c$, where $m_{\rm nf}$
is the number of free $M_{\rm n}$ and $c$ is the number of the complexes.
The total number of $E$, including both free and bound ones,
is denoted by $e_{\rm t}$.
The usual rate equation approach for the complex yields
$k_{1}m_{\rm nf}(e_{\rm t}-c)-(k_{2}+k_{3})c=0$ when quasiequilibrium is
established. Hence the mean complex number is given by,
\begin{equation}
c=e_{\rm t}\frac{m_{\rm nf}}{K_{\rm m}+m_{\rm nf}}.
\label{c-rate}
\end{equation}
On the other hand, through an exact analysis,
Levine and Hwa\cite{metabolicHwa}
obtained a modified expression in the limit $k_3\rightarrow 0$,
\begin{equation}
c=e_{\rm t} {m_{\rm n}\over K+m_{\rm n}},
\label{MM-c}
\end{equation}
where $K=K_{\rm m}+e_{\rm t}$.
The two expressions converge to the exact result
in both the linear regime $m_{\rm n}\ll K$ and the saturated regime
$m_{\rm n}\gg K$.
They differ only in the crossover regime $m_{\rm n}\simeq K$, where no exact
result is available in the general case, though either of the two can be used
as approximate expressions.

Using equation (\ref{MM-c}), we write the transport flux as,
\begin{equation}
\label{apprflux}v(m_{\rm n})=k_{3}c=\frac{v_{\rm max}m_{\rm n}}{K+m_{\rm n}},
\end{equation}
where $v_{\rm max}=k_{3}e_{\rm t}$.
Following the fast equilibration assumption, we may now describe
the mRNA export under the MM kinetics using the reduced description
(\ref{linear-transport-decay}) with an effective transport coefficient
$k=v_{\rm max}/(K+m_{\rm n})$ that decreases with increasing $m_{\rm n}$.
Again, taking the burst limit for the mRNA production,
we arrive at the following master equation for the joint distribution
of $m_{\rm n}$ and $m_{\rm c}$:
\begin{eqnarray}
\label{mastMM}
\frac{dP(m_{\rm n},m_{\rm c},t)}{dt}&=&\sum_{b=0}^{m_{\rm n}}\lambda
G(b)P(m_{\rm n}-b,m_{\rm c},t)
-\sum_{b=0}^{\infty}\lambda G(b)P(m_{\rm n},m_{\rm c},t)\nonumber\\
&&+(\varepsilon_{\rm n}\varepsilon_{\rm c}^{-1}-1)
v(m_{\rm n})P(m_{\rm n},m_{\rm c},t)\nonumber\\
&&+\delta(\varepsilon_c-1)m_{\rm c}P(m_{\rm n},m_{\rm c},t).
\end{eqnarray}
Since the nonlinear function $v(m_{\rm n})$ in the MM transport does
not allow for the closure
of equations for the second moments of the distribution,
approximate treatment of the master equation is necessary.

The two limiting situations $\langle m_{\rm n}\rangle\gg \langle b\rangle$
and $\langle m_{\rm n}\rangle\ll \langle b\rangle$ call for different
considerations. In the former case, the contribution of
each burst event on the total nuclear mRNA population is small,
so that perturbative treatment around the average
$\langle m_{\rm n}\rangle$ is appropriate. The latter case, however,
corresponds to the situation where the nuclear mRNA from each burst
event is essentially cleared before the next one arrives.
The two cases are treated separately below.

\subsubsection{Linear approximation}

We first consider the weak fluctuation case $\langle m_{\rm n}\rangle\gg
\langle b\rangle$. A general scheme to perform the noise calculations
is the van Kampen's $\Omega$-expansion whose lowest order terms
reproduce the macroscopic rate equations and the next order terms yield
a linear Fokker-Planck equation (FPE), which is often called the
linear noise approximation\cite{stochasticKampen}.
In Appendix A we
derive the noise strengths under the LNA. Here, we outline a more
direct yet equivalent way to obtain the results.

The approximation we introduce is to replace (\ref{apprflux}) with
its linear expansion at $m_{\rm n}=\langle m_{\rm n}\rangle$:
\begin{equation}
\label{MMflux}v\simeq k_{\rm eff}(m_{\rm n}+m_{0}).
\end{equation}
Here $k_{\rm eff}=v_{\rm max}K/(K+\langle m_{\rm n}\rangle)^{2}$ and
$m_{0}=\langle m_{\rm n}\rangle^{2}/K$. Substituting (\ref{MMflux}) into
(\ref{mastMM}), we may compute moments of the distribution
$P(m_{\rm n},m_{\rm c},t)$ in the same way as in Sec. 2.1.
In fact, equations (\ref{linear-flux})-(\ref{mc2apr}) remain
valid if we make the substitution $k\rightarrow k_{\rm eff}$ and
$m_{\rm n}\rightarrow m_{\rm n} + m_0$.
After rearranging the terms, we obtain:
\begin{eqnarray}
\label{mnmm}\frac{\sigma_{m_{\rm n}}^{2}}{\langle m_{\rm n}\rangle}
&=& (\frac{\langle m_{\rm n} \rangle}{K}+1) (\langle b\rangle+1),\\
\label{mcmm}\frac{\sigma_{m_{\rm c}}^{2}}{\langle m_{\rm c}\rangle}
&=&\langle b\rangle+1-\langle b\rangle\frac{\langle
m_{\rm n}\rangle}{\frac{K\langle m_{\rm c}\rangle}{K+\langle
m_{\rm n}\rangle}+\langle m_{\rm n}\rangle}.
\end{eqnarray}

\subsubsection{Independent burst approximation}

In the case $\langle m_{\rm n}\rangle\ll \langle b\rangle$, the mRNA
produced in a given burst has sufficient time to exit the nucleus
before the next burst arrives. It is then appropriate to consider
the independent burst approximation, where  individual burst
events contribute additively to $m_{\rm n}(t)$ and $m_{\rm c}(t)$:
\begin{eqnarray}
m_{\rm n}(t)&=&\sum_{t_i<t}\xi(b_i,t-t_i),\\
m_{\rm c}(t)&=&\sum_{t_i<t}\eta(b_i,t-t_i).
\label{burst}
\end{eqnarray}
Here $b_i$ is the size of the $i$th burst which takes place at $t_i$, and
$\xi(b,t)$ and $\eta(b,t)$ are the number of mRNAs in the nucleus and in the
cytoplasm, respectively, generated by a single burst of size $b$ at $t=0$.

Three independent stochastic processes contribute to the statistical
properties of the time series $m_{\rm n}(t)$ and $m_{\rm c}(t)$:
i) the time of the burst events $t_i$, which we assume to be
Poisson at a rate $\lambda$; ii) the size $b_i$
of individual bursts which follow the geometric distribution $G(b)$;
and iii) the stochastic nature of the MM transport and mRNA decay in
the small copy number regime. In the following discussion we shall
focus on the noise effects due to processes i) and ii), while
neglecting stochasticity in iii). The latter approximation is
justified by noting that the most significant contributions to the
quantities computed below are from the period when $\xi (b,t)$ and
$\eta (b,t)$ are large and their relative fluctuations are small.
Denoting by $x (b,t)=\langle\xi (b,t)\rangle$ and
$y(b,t)=\langle\eta (b,t)\rangle$, we obtain the following
expressions for the moments:
\begin{eqnarray}
\langle m_{\rm n}\rangle&=&\sum_b G(b)\int_0^\infty x(b,t)\lambda dt,
\label{x-int}\\
\langle m_{\rm n}^2\rangle&=&\langle m_{\rm n}\rangle^2
+\sum_b G(b)\int_0^\infty x^2(b,t)\lambda dt,\label{x2-int}\\
\langle m_{\rm c}\rangle&=&\sum_b G(b)\int_0^\infty y(b,t)\lambda dt,
\label{y-int}\\
\langle m_{\rm c}^2\rangle&=&\langle m_{\rm c}\rangle^2
+\sum_b G(b)\int_0^\infty y^2(b,t)\lambda dt,\label{y2-int}\\
\langle m_{\rm n}m_{\rm c}\rangle&=&\langle m_{\rm n}\rangle
\langle m_{\rm c}\rangle
+\sum_b G(b)\int_0^\infty x(b,t)y(b,t)\lambda dt.\label{xy-int}
\end{eqnarray}
Under the MM transport (\ref{apprflux}), the dynamical equations for
$x(b,t)$ and $y(b,t)$ are given by:
\begin{eqnarray}
{dx\over dt}&=&-v_{\rm max}{x\over K+x},\label{x-eq}\\
{dy\over dt}&=&v_{\rm max}{x\over K+x}-\delta y,
\label{y-eq}
\end{eqnarray}
with the initial condition $x(b,0)=b$ and $y(b,0)=0$.

As shown in Appendix B, the sum and integrals in equations (\ref{x-int})
and (\ref{x2-int}) can be worked out exactly to give:
\begin{eqnarray}
\langle m_{\rm n}\rangle&=&{\lambda\over v_{\rm max}}
\langle b\rangle\Bigl (\langle b\rangle +K +{1\over 2}\Bigr),\label{m_n_ind}\\
\sigma_{m_{\rm n}}^2
&\equiv&\langle m_{\rm n}^2\rangle-\langle m_{\rm n}\rangle^2
={\lambda\over v_{\rm max}}
\langle b\rangle\Bigl[2\langle b\rangle^2+(K+2)\langle b\rangle
+{K\over 2}+{1\over 3}\Bigr].
\label{sigma_n_ind}
\end{eqnarray}
Hence,
\begin{equation}
{\sigma_{m_{\rm n}}^2\over\langle m_{\rm n}\rangle}
=\langle b\rangle +{1\over 2}
+{\langle b\rangle^2+\langle b\rangle+{1\over 12}\over K
+{1\over 2}+\langle b\rangle}. \label{IBANsmn}
\end{equation}

The mean value of $m_{\rm c}$ can be obtained from flux balance, i.e.,
$\lambda\langle b\rangle=\delta\langle m_{\rm c}\rangle$ or
$\langle m_{\rm c}\rangle = (\lambda/\delta)\langle b\rangle$.
The calculation of $\langle m_{\rm c}^2\rangle$ is a bit more involved
which we relegate to Appendix B.
Assuming $\delta\ll v_{\rm max}$, i.e., the decay time of an mRNA
molecule is much longer than the fastest release time of one mRNA to
the cytoplasm, we may write the result in the form:
\begin{equation}
{\sigma_{m_{\rm c}}^2\over\langle m_{\rm c}\rangle}
=\Bigl (\langle b\rangle+{1\over 2}\Bigr)\Psi(u,w), \label{IBANsmc_a}
\end{equation}
where $u=\langle b\rangle\delta/v_{\rm max}$ and
$w=K\delta/v_{\rm max}$. The function $\Psi$ is given by
\begin{equation}
\Psi(u,w)=2\int_0^1dx\int_x^1 dx_1 {e^{w\ln (x/x_1)}\over [1+u (x_1-x)]^3}.
\label{psi-def}
\end{equation}
Since the integrand is less than 1, we have $\Psi(u,w)\leq 1$.

We have not managed to find a closed form expression for $\Psi(u,w)$,
but the integral can be worked out in the two limiting cases:
i) $u=0, \Psi(0,w)=1/(1+w)$; and
ii) $w=0, \Psi(u,0)=1/(1+u)$.
These two expressions also set upper bounds for $\Psi(u,w)$
in general. An approximate expression that is consistent with the two
limits and also verified by numerical integration of (\ref{psi-def})
is given by:
\begin{equation}
\Psi(u,w)\simeq {1\over 1+u+w}.
\label{psi-appro}
\end{equation}
In terms of $\langle m_{\rm n}\rangle$ and $\langle m_{\rm c}\rangle$
and with the help of (\ref{m_n_ind}) and the flux balance condition,
equation (\ref{IBANsmc_a}) can be rewritten as:
\begin{equation}
{\sigma_{m_{\rm c}}^2\over\langle m_{\rm c}\rangle}
\simeq
{\langle b\rangle+{1\over 2}\over
1+{\langle m_{\rm n}\rangle\over\langle m_{\rm c}\rangle}
{K+\langle b\rangle\over K+\langle b\rangle +{1\over 2}}}.
\label{IBANsmc}
\end{equation}

\section{Results and discussion}

Let us first summarize the analytical results derived in Section 2 when the
average burst size $\langle b\rangle\gg 1$. In general, noise strength of
mRNA copy number in the two compartments can be expressed in the form,
\begin{eqnarray}
{\sigma_{m_{\rm n}}^2\over\langle m_{\rm n}\rangle}
&=&\alpha \langle b\rangle+1,\label{alpha-def}\\
{\sigma_{m_{\rm c}}^2\over\langle m_{\rm c}\rangle}
&=&{\langle b\rangle\over 1
+\beta\langle m_{\rm n}\rangle/\langle m_{\rm c}\rangle}+1.
\label{beta-def}
\end{eqnarray}
For the linear model, we have $\alpha=\beta=1$. In the MM case
where the transporter may become the bottleneck in the process,
$\alpha=\beta=1+\langle m_{\rm n}\rangle/K$ if the slow transport leads
to the nuclear accumulation of mRNA, i.e.,
$\langle m_{\rm n}\rangle\gg\langle b\rangle$.
In the opposite limit
$\langle m_{\rm n}\rangle\ll\langle b\rangle$, where there is nuclear
clearance between successive burst events,
$\alpha\simeq1+\langle b\rangle/(K+\langle b\rangle)$
and $\beta\simeq 1$.

The general trend of noise attenuation on the cytoplasmic mRNA
due to delay in nuclear transport is now clear.
Retention of the mRNA inside the nucleus decreases
the effective burst size and hence the noise strength of the cytoplasmic mRNA.
Even at a fixed ratio of $\langle m_{\rm n}\rangle$ to
$\langle m_{\rm c}\rangle$, further reduction of the noise is
possible in the nonlinear MM transport when $\langle m_{\rm n}\rangle$
is greater than both the dissociation constant $K$ and burst size
$\langle b\rangle$.
On the other hand, the independent burst approximation yields a
$\beta$ value close to one, extending the validity of the linear model
when the noise strength of $m_{\rm c}$
is considered as a function of the ratio of mean copy numbers
$\langle m_{\rm n}\rangle/\langle m_{\rm c}\rangle$.

Fluctuations in the nuclear mRNA level, on the other hand, exhibits
a somewhat different behavior. The parameter $\alpha$ that
characterizes the noise strength of $m_{\rm n}$ reaches the minimum value
1 in the linear model, but increases when the saturation effect in
the MM transport kicks in. Thus queuing results in an enhanced
fluctuation upstream of the transport channel.

To check the accuracy of the analytic results under
parameter values that broadly correspond to the mammalian cell
gene expression experiments mentioned above, we have carried out
simulations of the stochastic MM transport defined by (\ref{MM-rates}),
following the Gillespie's exact algorithm\cite{Gillespie}.
The unit of time is chosen such that the mRNA decay rate $\delta=1$.
The number of transport channels is set to $e_{\rm t}=10$.
We fix the average mRNA production rate $\lambda\langle b\rangle$
relative to the mRNA decay rate $\delta$
such that the mean cytoplasmic mRNA copy number
$\langle m_{\rm c}\rangle=\lambda\langle b\rangle/\delta=40$.
For easy comparison with the experimental measurements and analytic
results, we also fix the mean nuclear mRNA copy number
$\langle m_{\rm n}\rangle=20$. With these parameter values,
the noise strengths in the linear model are given by
$\sigma_{m_{\rm n}}^2/\langle m_{\rm n}\rangle=\langle b\rangle +1$
and
$\sigma_{m_{\rm c}}^2/\langle m_{\rm c}\rangle={2\over 3}\langle b\rangle+1$,
respectively.
Simulations are then performed to examine the effect of channel saturation
on the noise strengths by varying the MM parameters
$k_{1}$ and $k_{3}$ in such a way that
the mean nuclear mRNA copy number stays at the value set above.
The unbinding rate of the mRNA-transporter complex $EM_{\rm n}$
is fixed at a low value $k_2=0.1$.

\begin{figure}
\includegraphics[width=5in]{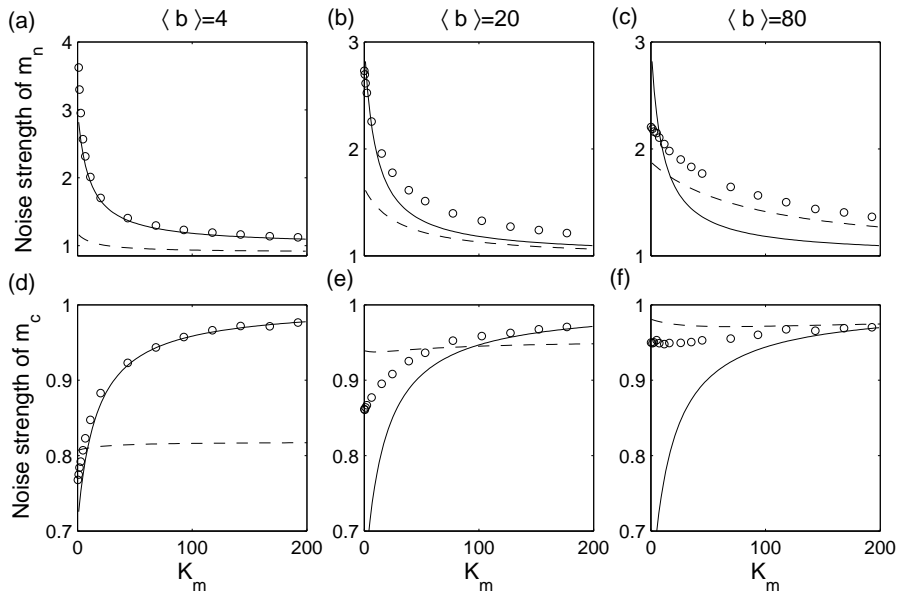}
\caption{Noise strengths (normalized by their corresponding values in
the linear model) of the nuclear (upper panel) and
the cytoplasmic (lower panel) mRNA copy numbers against
the Michaelis constant $K_{\rm m}$ which controls saturation of
transport channels. Here $e_{\rm t}=10$, $\langle m_{\rm n}\rangle=20$
and $\langle m_{\rm c}\rangle=40$.
Results are shown for stochastic simulation (open circles),
LNA (solid line) and IBA (dashed line).
The overall noise strength is set by the mean burst size
$\langle b\rangle$ which is chosen to be (in units of mRNA copy number):
4 [(a) and (d)]; 20 [(b) and (e)]; 80 [(c) and (f)]. See text
for the choice of other model parameters.
} \label{noisestrength}
\end{figure}

\Fref{noisestrength} shows a comparison of simulation data (open circles)
and analytic results under the LNA (solid line) and the IBA (dashed line),
respectively. Plots on the upper panel give the noise strength of nuclear
mRNA as a function of the Michaelis constant $K_{\rm m}=(k_2+k_3)/k_1$.
Fluctuations in $m_{\rm n}$ grow as the saturation
effect becomes more prominent on the low $K_{\rm m}$ side.
An opposite trend is seen in the fluctuations of the cytoplasmic
mRNA copy number $m_{\rm c}$ shown in plots on the lower panel.
As expected, the LNA results (solid line) agree well with the
simulation data (circles) in the weak burst regime
[(a)and (d)], in which case the overall noise strength
(as compared to the mean mRNA copy number) is weak.
On the other hand, the IBA results (dashed line) represent a better
approximation in the strong burst regime [(c)and (f)]\cite{note2}.
Therefore each of the two approximations perform reasonably well
in their respective regimes of validity, and are complementary to each other.

\begin{figure}
\includegraphics[width=5in]{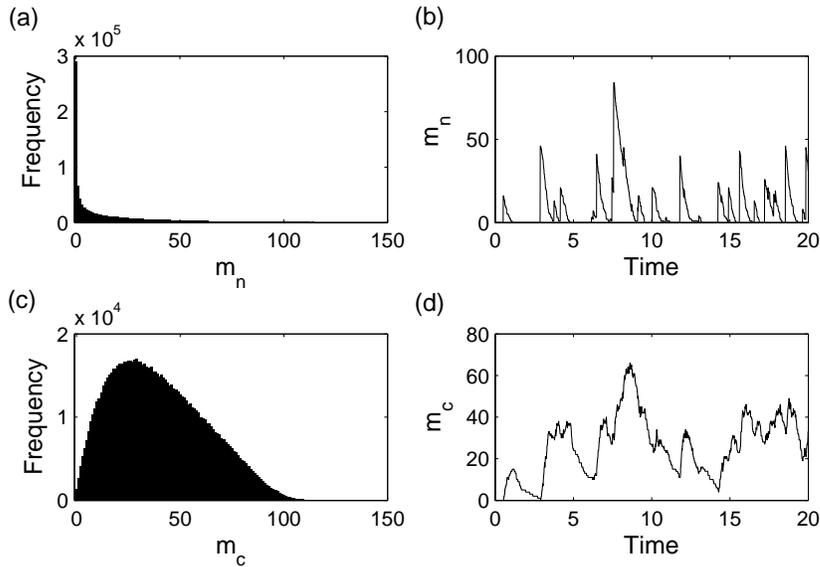}
\caption{Distributions of mRNA copy numbers in the nucleus (a) and in the
cytoplasm (c), obtained from 500 000 repeated runs of the Gillespie exact
simulation. Sample time courses of $m_n$ and $m_c$ in a single simulation
are shown in (b) and (d).
Parameter values are: $\lambda=2$, $\langle b\rangle=20$,
$k_{1}=1.56$, $k_{2}=0.1$, $k_{3}=10$, $e_{\rm t}=10$ and $\delta=1$.}
\label{PdfTimeCourse}
\end{figure}

\Fref{PdfTimeCourse} shows the actual distribution [(a) and (c)] and
sample time course [(b) and (d)] of the nuclear and cytoplasmic mRNA
copy number, respectively, generated from simulations at $K_{\rm m}=6.5$.
Other model parameters are the same as that of \fref{noisestrength} (b)
and (e), which represents a borderline case for the two approximate
treatments. It is seen from \fref{PdfTimeCourse} (b) that
$m_{\rm n}$ falls below $K=16.5$ most of the time,
so approximating the MM transport by a linear expansion at
$m_{\rm n}=\langle m_{\rm n}\rangle$ is too crude.
On the other hand, there are occasional overlaps of the mRNA produced
in successive bursts. This has the effect of slowing down the mRNA
transport than what is assumed in the IBA, leading to a somewhat
lower $m_{\rm c}$ noise as seen in the left part of
\fref{noisestrength} (e).

\begin{figure}
\includegraphics[width=5in]{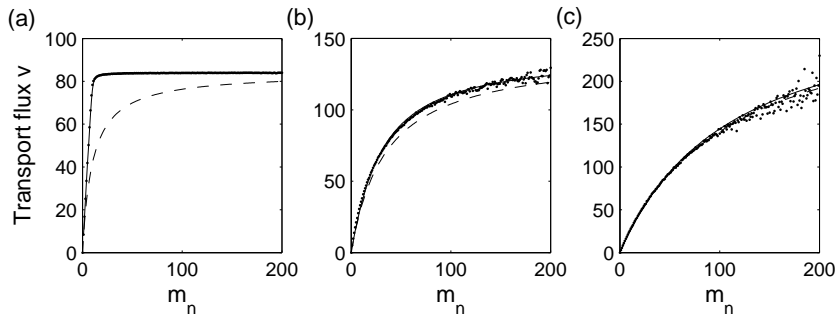}
\caption{Transport flux as calculated from the stochastic simulations (dots),
equation \eref{c-rate} (solid line) and equation \eref{apprflux}
(dashed line). The transport parameters are: (a) $k_3=8.4$,
$K_{\rm m}=0.1$;
(b) $k_3=14$, $K_{\rm m}=24.5$; (c) $k_3=30$, $K_{\rm m}=102.5$.
Other parameter values are given by $\lambda=2$,
$\langle b\rangle=20$, $k_{2}=0.1$, $e_{\rm t}=10$ and $\delta=1$.
Note that $\langle m_{\rm n}\rangle=20$ and $\langle m_{\rm c}\rangle=40$
are the same as before.}
\label{checkMM}
\end{figure}

We have also examined the validity of equation \eref{apprflux}
for the mean transport flux. At a given number $c$ of complexes $EM_{\rm n}$,
the expected mRNA export flux is $k_3 c$. Therefore $v(m_n)$ can
be obtained in the simulations from the conditional average of $c$
at a given $m_n$. As shown in \fref{checkMM}, equation \eref{apprflux}
(dashed line) fits the simulation data (dots) very well
when $K_{\rm m}\geq\langle m_{\rm n}\rangle$ [(c)],
but quite poorly in the regime $K_{\rm m}\ll\langle m_{\rm n}\rangle$ [(a)].
Surprisingly, the classic MM equation \eref{c-rate}, suitably
modified to be considered as a function of $m_n$, agrees with the numerical
result extremely well for both large and small values of $K_{\rm m}$.
Note that equation \eref{apprflux} was derived under the quasi-equilibrium
approximation $k_3\ll k_1, k_2$ which does not hold in the present case.
This inaccuracy may also contribute to the discrepancy between
our analytic results and simulation data at small $K_{\rm m}$ in
\fref{noisestrength}.

\section{Conclusions and outlook}

The main conclusion of our work is that the nuclear envelope, which
sets a natural barrier for the exodus of mature mRNA to the cytoplasm
in an eukaryotic cell, can significantly attenuate the effect of
transcriptional bursting on the downstream protein population.
The extent of the noise reduction on the cytoplasmic mRNA copy number
is controlled by the transport efficiency. A high transport rate
has a weak effect in noise reduction and essentially brings one
back to the one-compartment model considered previously.
On the other hand, a low transport rate turns the nucleus into a
buffer for the bursting noise, thereby reducing the temporal
variation of the mRNA copy number in the cytoplasm.
This effect is more dramatic in the saturated regime under the
Michaelis-Menten dynamics where, due to the limited availability of
transport channels, the mRNA export becomes a Poisson process
unaffected by the bursty input.

Our results can be used to re-estimate the mean mRNA burst size
in the experiment by Raj and his colleagues.
Take the linear transport as an exmaple, the noise
strength of the total mRNA copy number in the cell can be obtained from
equations \eref{mn2apr}, \eref{mnmcapr} and \eref{mc2apr}:
\begin{equation}
{\sigma_{m_{\rm n}+m_{\rm c}}^2\over\langle m_{\rm n}+m_{\rm c}\rangle}
=\langle b\rangle+1+\langle b\rangle
{\langle m_{\rm n}\rangle\langle m_{\rm c}\rangle
\over(\langle m_{\rm n}\rangle+\langle m_{\rm c}\rangle)^2}.
\label{mRNA-total}
\end{equation}
Thus the average burst size $\langle b\rangle$ can be obtained from
the measured total mRNA fluctuations and the ratio
$\langle m_n\rangle/\langle m_c\rangle$
of nuclear to cytoplasmic mRNA.
If $30\%$ of the mRNA accumulate in the nucleus, the burst size should
be $83\%$ of that estimated from the model without transport.
With the help of the two-compartment models,
it would be interesting to revisit the single-molecule
experiment data of Raj {\it et al.} to gain a more complete view of the
role of transport on the characteristics of the mRNA noise
generated by transcriptional bursting.

Previous studies suggest that for eukaryotes, such as
{\it S. cerevisiae}\cite{eukaryotic1Oshea},
{\it Dictyostelium}\cite{DictyosteliumSinger} and
mammalian cells\cite{mamalianTyagi}, transcriptional bursting is a
dominant source of the noise from the internal molecular circuits
known as the ``internal noise''. It is interesting to note that
the stress-related genes, which demand a fast response
and an accelerated export rate, are noisier than the essential
house-keeping genes such as proteasome genes\cite{naturalnoiseBarKai}.
In our limited transport capacity model described by the MM dynamics,
the downstream noise is indeed stronger in the linear regime and weaker
in the saturated regime. The linear regime also allows for a faster
response to external stimulus as there is no queuing effect.
In this respect, one can not help but wonder whether special processing
and export channels are in place in the nucleus for the fast
release of stress response genes without congestion.

Finally, we would like to mention that
noise propagation through capacity-limited channels with a bursty input
is a general phenomenon not limited to the transcriptional bursting,
and hence the discussions initiated here can be of broader significance.
A somewhat peculiar behavior associated with the
limited-capacity transport is that, as the extent of channel saturation
increases while maintaining the same average transport current, 
the noise level of the upstream population increases while
that of the downstream population decreases (see \fref{noisestrength}). 
This counterintuitive phenomenon is
neither stochastic resonance (SR) nor stochastic focusing (SF) proposed by
Paulsson \etal\cite{focusingPNASEhrenberg,focusingPRLEhrenberg}.
SR is usually
related to periodic signal detection, where the signal noise is
typically external, while SF exploits signal noise to make a gradual
response mechanism work more like a threshold mechanism. Here, the
interesting phenomenon we observe is due to a quite different
mechanism: the degradation of upstream species and production of
downstream species share a common reaction (here transport), whose
effective order at steady state can be tuned from zero to one by
a combination of reaction parameters.

\section*{Acknowledgments}
We thank HG Liu and XQ Shi for helpful discussions. LPX would like to
thank the Physics Department, Hong Kong Baptist University for hospitality
where part of the work was carried out. This work was
supported by the National Natural Science Foundation of China under
grant 10629401, and by the Research Grants Council of the HKSAR under grant
HKBU 2016/06P.

\appendix\setcounter{section}{0}
\section{The $\Omega$-expansion}
Equations (\ref{mnmm}) and (\ref{mcmm}) can be equivalently derived
using the more formal $\Omega$-expansion which applies when
fluctuations of $m_{\rm n}$ and $m_{\rm c}$ are weak. ($\Omega$ here stands for
the system volume.) To set the notation straight, the rate equation
for the copy number $x_i$ of molecule $i$ (i.e., $m_{\rm n}$ or $m_{\rm c}$ in
the present case) is given by,
\begin{equation}
dx_i/dt =\sum_q S_{iq}V_q= (\textbf{S}\cdot\textbf{V})_i,
\label{rate-eq-x}
\end{equation}
where $S_{iq}$ is the stoichiometric coefficient of molecule $i$ in
reaction $q$, and $V_q$ (which depends on the copy numbers in
general) is the propensity of reaction $q$. Stochasticity in the
problem is assumed to be due to fluctuations in the propensity $V_q$
with a strength set at $V_q^{1/2}$ due to the underlying Poisson
process.

In the large volume limit and for the steady state, the joint distribution
of the $x_i$'s, which satisfies the Fokker-Planck equation,
can be approximated by a gaussian centered at the solution to
the equation $d\textbf{x}/dt=\textbf{S}\cdot\textbf{V}=0$.
The width of the distribution is parametrized by a covariance
matrix $\textbf{C}$ which satisfies the equation \cite{lnaEhrenberg}:
\begin{equation}
\label{fdt} \textbf{AC}+\textbf{CA}^{T}+\textbf{B}=\textbf{0}.
\end{equation}
Here
\begin{eqnarray}
\label{dift}\textbf{A}_{ij}
&=&\frac{\partial(\textbf{S}\cdot \textbf{V})_{i}}{\partial x_{j}},\\
\label{diffusion} \textbf{B}_{ij}&=&\sum_q V_q S_{iq}S_{jq},
\end{eqnarray}
all evaluated at the steady state. Specializing on our two-component problem,
there are three reactions: the bursting reaction leading to the production
of $m_{\rm n}$, the MM transport reaction for nuclear export, and the
mRNA decay in the cytoplasm. Simple calculations yield:
\begin{eqnarray} \label{A}
\textbf{A}=\left (\begin{array}{cc}-\frac{v_{\rm max}K}{(K+\langle
m_{\rm n} \rangle)^{2}} & 0\\ \\\frac{v_{\rm max}K}{(K+\langle m_{\rm n}
\rangle)^{2}} & -\delta
\end{array}\right).
\end{eqnarray}
Application of (\ref{diffusion}) yields:
\begin{eqnarray}
\label{B11}
B_{11}&=&\frac{v_{\rm max}\langle m_{\rm n}
\rangle}{K+\langle m_{\rm n} \rangle}+\sum_{b}\lambda G(b)b^{2}\nonumber\\
&=&\frac{v_{\rm max}\langle m_{\rm n} \rangle}{K+\langle m_{\rm n} \rangle}
+\lambda \langle b^{2}\rangle\nonumber\\
&=&2\lambda (\langle b\rangle+\langle b\rangle^{2}).
\end{eqnarray}
In the last step, we have used the flux-balance condition and
$\langle b^{2}\rangle=2\langle b\rangle^{2}+\langle b\rangle$.
Other matrix elements of $\textbf{B}$ can be obtained as:
\begin{eqnarray}
\label{Bothers} B_{12}=B_{21}=-\lambda \langle b\rangle, \quad
B_{22}=2\lambda \langle b\rangle.
\end{eqnarray}
Substituting the values of $\textbf{A}$ and $\textbf{B}$ into
equation \eref{fdt} gives:
\begin{eqnarray}
\label{C11} \sigma_{m_{\rm n}}^2
&=&C_{11}=\frac{\langle m_{\rm n} \rangle (\langle m_{\rm n}
\rangle+K)}{K} (\langle
b\rangle+1),\\
\label{C12} \langle m_{\rm n} m_{\rm c}\rangle-\langle m_{\rm n}\rangle\langle
m_{\rm c}\rangle&=&C_{12}=C_{21}=\langle b\rangle\frac{\langle
m_{\rm n}\rangle \langle m_{\rm c} \rangle}
{\frac{K\langle m_{\rm c}\rangle}{K+\langle m_{\rm n} \rangle}
+\langle m_{\rm n}\rangle},\\
\label{C22} \sigma_{m_{\rm c}}^2
&=&C_{22}=\langle m_{\rm c} \rangle (\langle b\rangle+1)
-\langle b\rangle\frac{\langle m_{\rm n} \rangle \langle m_{\rm c}
\rangle}{\frac{K\langle m_{\rm c} \rangle}{K+\langle m_{\rm n}
\rangle}+\langle m_{\rm n}\rangle},
\end{eqnarray}
from which equations (\ref{mnmm}) and (\ref{mcmm}) follow.

\section{Integrals in the independent burst approximation}

A convenient way to carry out the integrals in (\ref{x-int})-(\ref{y2-int})
is to convert them into integration over $x$, which decreases monotonically
from its initial value $b$ to 0 in a single burst event,
with the help of (\ref{x-eq}) and (\ref{y-eq}).
Following this procedure, we may write,
\begin{eqnarray}
\int_0^\infty dt x(b,t)&=&-\int_0^b {xdx\over dx/dt}
=\int_0^b dx{K+x\over v_{\rm max}}
={1\over v_{\rm max}} (K b+{1\over 2}b^2).\\
\int_0^\infty dt x^2 (b,t)&=&-\int_0^b {x^2dx\over dx/dt}
=\int_0^b xdx{K+x\over v_{\rm max}}\nonumber\\
&=&{1\over v_{\rm max}} ({1\over 2}Kb^2+{1\over 3}b^3).
\end{eqnarray}
To perform the averaging over $b$, we make use of
the following results for the geometric distribution,
\begin{eqnarray}
\langle b^2\rangle&=&\langle b\rangle (1+2\langle b\rangle),\nonumber\\
\langle b^3\rangle
&=&\langle b\rangle (1+6\langle b\rangle+6\langle b\rangle^2).\nonumber
\end{eqnarray}
With the help of these results, (\ref{m_n_ind}) are (\ref{sigma_n_ind})
are readily obtained.

The dependence of $y$ on $x$ follows the equation,
\begin{equation}
{dy\over dx}={dy/dt\over dx/dt}=-1+{\delta\over v_{\rm max}} (1+{K\over x})y,
\label{y-rate}
\end{equation}
which can be integrated to give,
\begin{equation}
y (x)=\int_x^b dx_1e^{\delta (x-x_1)/v_{\rm max}+w\ln (x/x_1)},
\label{y-x}
\end{equation}
where $w\equiv \delta K/v_{\rm max}$.

As a check, let us first consider
\begin{equation}
\int_0^\infty dt y=\int_0^b dx{K+x\over v_{\rm max}x}y.
\end{equation}
Using equation (\ref{y-rate}) and noting that $y (x=b)=y (x=0)=0$, we obtain,
\begin{equation}
\int_0^b dx{K+x\over v_{\rm max}x}y
=\delta^{-1}\int_0^b dx (1+{dy\over dx})=b/\delta.
\end{equation}
Hence,
\begin{equation}
\langle m_{\rm c}\rangle=\lambda \langle b\rangle/\delta,
\end{equation}
which is nothing but the conservation law.

We now consider
\begin{equation}
\int_0^\infty dt y^2=\int_0^b y\delta^{-1}({dy\over dx}+1)dx
=\delta^{-1}\int_0^b ydx.
\label{y2A1}
\end{equation}
Using equation (\ref{y-x}) and perform the substitution $x\rightarrow bx$,
we obtain,
\begin{equation}
\int_0^\infty dt y^2={b^2\over\delta}
\int_0^1dx\int_x^1 dx_1 e^{b (\delta/v_{\rm max}) (x-x_1)+w\ln (x/x_1)}.
\end{equation}

The averaging over $b$ can now be readily carried out. Using the result
$\sum_{b=0}^\infty b^2a^b=a (1+a)/(1-a)^3$, we obtain,
\begin{equation}
\sum_b G(b) \int_0^\infty dt y^2 (b,t)
=\delta^{-1}\int_0^1dx\int_x^1 dx_1 {e^{w\ln (x/x_1)}\over 1
+\langle b\rangle}{a (1+a)\over (1-a)^3}.
\end{equation}
Here $a=\langle b\rangle e^{(x-x_1)\delta/v_{\rm max}}/(1+\langle b\rangle)$.

To avoid run-away accumulation of mRNAs in the nucleus, we require
$\lambda\langle b\rangle<v_{\rm max}$.
Therefore $\langle m_{\rm c}\rangle <v_{\rm max}/\delta$.
Note that $\langle m_{\rm c}\rangle>1$ automatically implies
$\delta/v_{\rm max}$ to be a small quantity.
In this case, we can approximate
$a\simeq \langle b\rangle [1+ (x-x_1)\delta/v_{\rm max}]/(1+\langle b\rangle)$.
Consequently,
\begin{equation}
\sum_b G(b) \int_0^\infty dt y^2 (b,t)
={\langle b\rangle\bigl(\langle b\rangle+{1\over 2}\bigr)\over\delta}\Psi(u,w),
\label{y2A2}
\end{equation}
where $u=\langle b\rangle\delta/v_{\rm max}$ and $\Psi(u,w)$ is
given by \eref{psi-def}.

Finally, the integral in equation \eref{xy-int} can be rewritten in the form,
\begin{equation}
\int_0^\infty dt x(b,t)y(b,t)=
\int_0^b x\delta^{-1} ({dy\over dx}+1)dx
={b^2\over 2\delta}-\delta^{-1}\int_0^b ydx.\label{xyA1}
\end{equation}
Comparing with \eref{y2A1} and using \eref{y2A2}, we obtain,
\begin{equation}
\sum_b G(b) \int_0^\infty dt x(b,t)y(b,t)
={\langle b\rangle\bigl(\langle b\rangle+{1\over 2}\bigr)\over\delta}
\bigl[1-\Psi(u,w)\bigr].
\label{xyA2}
\end{equation}


\begin{thebibliography}{10}
\bibitem{note1} The study by Vargas and colleagues on
the diffusion of mRNA particles in the nucleus of Chinese hamster
ovary (CHO) cell, the same system used in Ref. \cite{mamalianTyagi},
obtained a diffusion constant $D=0.03-0.06\mu$m$^2$/sec
\cite{transport2Tyagi}. Given that the diameter of CHO nucleus is
about 5 $\mu$m, the time needed for the mRNP complexes to disperse
throughout the nucleus is of the order of a few minutes, much
shorter than the nuclear dwelling time of several hours.
\bibitem{note2} The dashed lines in the lower panel of
\fref{noisestrength} are obtained by evaluating
the integral \eref{psi-def} numerically. Direct application of
\eref{IBANsmc} yields nearly identical results.
\end{thebibliography}
\end{document}